\documentclass[review,12pt]{elsarticle}

\usepackage{pgfplotstable} 

\usepackage{makecell}
\usepackage{xfrac}
\usepackage{siunitx,booktabs}
\usepackage{amsmath}
\usepackage{hyphenat}
\usepackage[english]{babel}

\usepackage{tabularx}

\usepackage{ragged2e}
\usepackage{longtable}
\usepackage[utf8]{inputenc}
\DeclareUnicodeCharacter{2212}{-}

\usepackage{array}
\usepackage[longtable]{multirow}
\usepackage{longtable}
\usepackage{tabularray}

\usepackage{graphicx}
\graphicspath{ {pics/} }
\usepackage{caption}
\usepackage{float}
\usepackage{graphicx, caption, subcaption}
\usepackage{amssymb}
\usepackage[utf8]{inputenc}
\DeclareUnicodeCharacter{2212}{-}

\usepackage{graphicx}
\journal{Photonics \& Nanostructures-Fundamentals and Appilications }
\pgfplotsset{compat=1.18}
\begin{document}

\begin{frontmatter}


\title{Design and Analysis of Arc-Shaped Single Core Photonic Crystal Fiber Sensor Based on Surface Plasmon Resonance}

\author[inst1]{Tasmiah Tunazzina}
\affiliation[inst1]{organization={Department of Electrical and Electronic Engineering},
            addressline={Chittagong University of Engineering and Technology}, 
            city={Chittagong},
            postcode={4349}, 
            country={Bangladesh}}

\author[inst1]{Fairuz Areefin Khan}
\author[inst1]{Anuva Chowdhury}


\begin{abstract}
The selection of a suitable plasmonic material is extremely crucial for achieving high-performance photonic crystal fiber-based surface plasmon resonance (PCF-SPR) sensors. However, most numerical investigations to date are limited to PCF-SPR sensors with conventional circularly coated plasmonic metals due to their availability and rigid properties. In this work, a single-core arc-shaped photonic crystal fiber is designed and studied with sensing ingredients coated outside the fiber. The simulation and numerical analyses were performed using the full-vector finite element method (FV-FEM) to examine the effects of using gold as active metal and also the deposition of $Ta_2 O_5$ on gold. The results show that the arc-shaped sensor with gold film can obtain the maximum wavelength interrogation sensitivity (WIS) of 9500 nm/RIU within the refractive index (RI) range of 1.25 to 1.39 while the maximum amplitude interrogation sensitivity (AIS) is reached 579.26 $RIU^{-1}$ at 1.39 and resolution is found to be $1.05\times{10^{-5}}$. However, depositing $Ta_2 O_5$ on the gold gives an improved maximum WIS and AIS of 22000 nm/RIU, and 1209.21 $RIU^{-1}$, respectively. With the coating of $Ta_2 O_5$, the resolution improves to $4.55\times{10^{-6}}$ making the proposed sensor design undoubtedly effective in detecting different biological samples with a wide range of RI. 
\end{abstract}



\end{frontmatter}


\section{Introduction}
\label{sec:sample1}
The surface plasmon resonance (SPR) sensor has gained significant attention in recent years as a valuable research topic. During the development of surface plasmon resonance-based sensors, fiber optic sensors have become a significant matter of interest due to their remote accessibility, high sensitivity, real-time detection, and compatibility with long-range sensing. Researchers have proposed fiber optic sensing based on fiber Bragg grating, single and multimode fiber, and micro-structured fiber based on internal and external sensing schemes \cite{app9050949},\cite{7222411}. As a distinctive and unique class of optical fibers, photonic crystal fiber (PCF) has become popular in the SPR-based sensing scheme as PCFs have various advantages regarding their flexibility in geometry, immunity to any electromagnetic interference, unusual optical properties, controllable birefringence, and high confinement compared to conventional optical fibers \cite{DAS2021100904}. Moreover, the study of using various highly-sensitive plasmonic materials for more stable, simple, and reliable performance has motivated the search for novel plasmonic materials and compatible PCF geometry.\\
Till now, the sensing mechanism of PCF-SPR sensors is classified into internal sensing scheme (ISS) and external sensing scheme (ESS) \cite{RIFAT2017311}. Obtaining accurate results from the internal sensing scheme is challenging due to the plasmonic materials and sensing samples coating the micron-sized air holes \cite{ARUNAGANDHI2019102590}. This complicates obtaining a constant metal coating and accuracy, which leads to adopting ESS, as in this scheme, plasmonic materials are placed on the outer surface \cite{SHAFKAT2020100324}. In PCF SPR sensors, the direct metal coating is widely recognized as more flexibility can be obtained through it. Different plasmonic materials, gold (Au), aluminum (Al), Titanium dioxide ($TiO_2$), silver (Ag), Copper (Cu), etc., have been used in research over the past few years. When the surface field is exposed, this examination of the practical outcomes of PCF with significant-performing plasmonic materials is considered. Among the two leading plasmonic materials, Silver (Ag) and gold (Au), gold (Au) is chemically reliable and bio-compatible. A peak resonance is also advanced in gold (Au)\cite{PMID:24755592}, whereas a lower peak value is found in silver (Ag). Besides, the analyte detection accuracy of gold could be more impressive as silver can be oxidized and reformulated easily. In the article \cite{s19173666}, the author proposed a gold (Au) and graphene layer sensor. After the graphene coating, an output of improved wavelength sensitivity has been found \cite{Kravets2014GrapheneprotectedCA},\cite{USHA2016986}. \\
Recently, the implementation of $Ta_2 O_5$ as the plasmonic material has been introduced because of its low waveguide loss in the near-infrared (NIR) range, comprehensive optical transparency, high refractive index, and environmental stability. These properties make it suitable for surface plasmon resonance (SPR) and evanescent field-based sensing \cite{Ezhilvalavan_1999}. Melwin et al. proposed an arc-shaped gold-coated PCF to achieve a maximum wavelength sensitivity of 14100 nm/RIU for the sensing range of 1.32-1.37 \cite{Melwin2023}.\\
This research holds significant importance as it addresses a gap in the existing literature by exploring the potential of the arc-shaped PCF SPR sensor for multi-analyte detection \cite{Melwin2023}. While gold has demonstrated effective performance as a plasmonic material, adding a $Ta_2 O_5$ coating enhances its wavelength sensitivity for multi-analyte detection \cite{9794733}. By utilizing different plasmonic materials, including gold and gold with $Ta_2 O_5$ coating, and increasing the number of air holes in the design, we aim to enhance the sensor's performance in terms of wavelength sensitivity, amplitude sensitivity, and resolution.\\
In this paper, utilizing the external sensing approach, we propose a PCF-SPR sensor with gold as active plasmonic material and $Ta_2 O_5$ as an overlayer to avoid oxidation and enhance sensitivity. Unlike the conventional full circular coating, we focus on detecting a wide range of analytes while minimizing the reliance on plasmonic materials by implementing the arc-shaped PCF-SPR sensor. The proposed sensor is designed and simulated using Finite Element Method (FEM) based COMSOL Multiphysics software. Although numerous PCF-based plasmonic sensors incorporated with $Ta_2 O_5$ have been designed and studied, it remains a potential research area.

\section{Structural Design and Theoretical Background}

A 3D schematic of the proposed photonic crystal fiber sensor is depicted in Fig.\ref{fig:Picture7}. It is composed of two hexagonal lattice rings of air holes with six analogous diameters $d = 1 \mu$m in the 1st loop and ten air holes in the 2nd loop with diameter $d_1 = 1.6 \mu$m. The air hole diameter of the outer lattice is made bigger to confine the light in the core region. Two air holes in the 2nd loop were intentionally removed to introduce better coupling between the evanescent wave and surface plasmon of the metal-analyte surface to increase the leakage of core-guided mode. Furthermore, an arc-shaped Au layer of thickness, ${t_A}_u$= 30 nm with length taken regarding arc angle, L = $60^\circ$, has been coated externally. The gold layer is coated in an arc shape covering the top three air holes so that light from the core region propagates through the cladding to interact with the gold layer. An overlayer of $Ta_2 O_5$ of thickness 10 nm is deposited over Au, which is kept in direct contact with the analyte, as clearly shown in Fig.\ref{fig:Picture7}. The diameter of the PCF is 11 $\mu$m with a 3 $\mu$m thickness of the analyte layer. To absorb the energy of outward scattering radiations, a 16 $\mu$m diameter perfectly matched layer (PML) has been incorporated.
Silica is chosen as background material, and its refractive index is calculated using Sellmeier equation \cite{Kitamura:07}
\begin{equation}
    {n_{silica}}^2 = 1 + \frac{A_1\lambda^2}{\lambda^2-B_1} + \frac{A_2\lambda^2}{\lambda^2-B_2} + \frac{A_3\lambda^2}{\lambda^2-B_3}
\end{equation}
    
where $A_1$=0.6961663, $B_1$=0.0684043, $A_2$=0.4079426, $B_2$=0.1162414,\\ 
$A_3$=0.8974794, $B_3$=9.896161,$ \lambda$ is the wavelength of the incident light and $n_{silica}$ is the wavelength-dependent refractive index of silica.

\begin{center}
    \centering
    \includegraphics[width=0.8\textwidth]{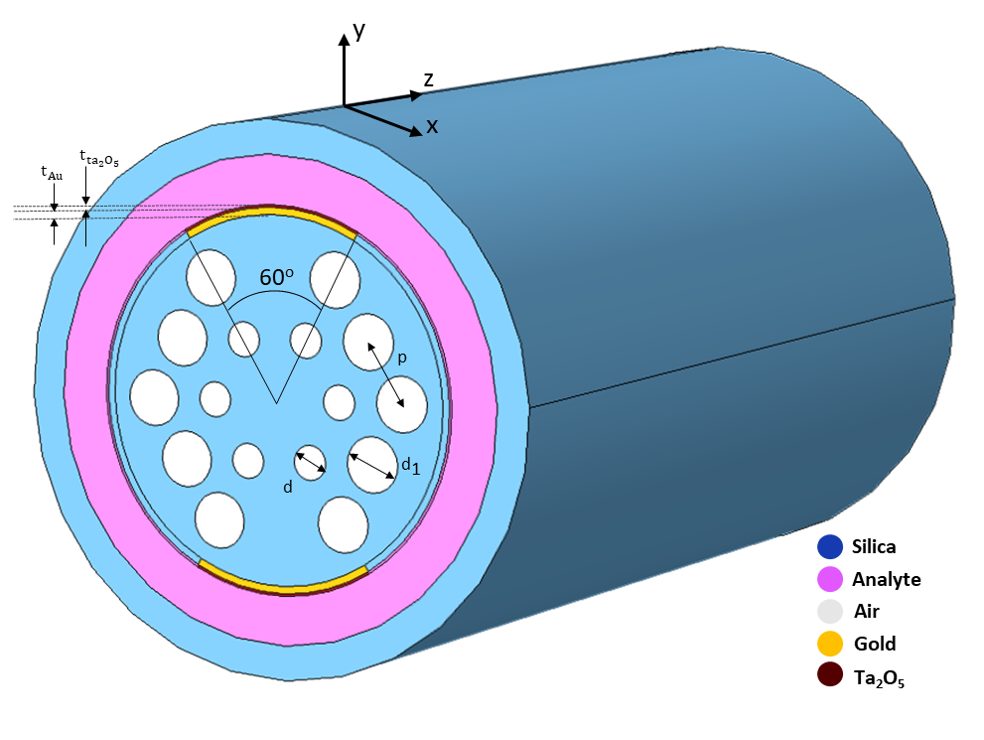}
    \captionof{figure}{3D Schematic of proposed arc-shaped gold coated PCF-SPR sensor.}
    \label{fig:Picture7}
\end{center}
\vspace{2pt}
\subsection{SPR phenomemon}
In the bulk assembly of negative electrons and equally charged positive ions, a longitudinal oscillation known as plasma oscillation is observed in the presence of an external electric field that causes the negative free electrons to attract toward positive ions giving rise to an attraction force and, consequently, a restoring force due to coulomb repulsion. In a metal-dielectric interface, the quantum of this plasma oscillation is referred to as surface plasmon or surface plasmon wave and is accompanied by a guided TM or p-polarized electric field, which is concentrated around the metal-dielectric interface. Solving Maxwell's equations, the dispersion relation of surface plasmon polaritons (or surface plasmons) which relates to the wavenumber ($K_{sp}$) of surface plasmon wave to the frequency of incident light $\omega$ can be found as \cite{inbook}
\begin{equation}\label{maxwell}
    K_{sp} = \frac{\omega}{c} \sqrt{\frac{\epsilon_m\epsilon_s}{\epsilon_m+\epsilon_s}}
\end{equation}
where $\epsilon_m$ and $\epsilon_s$ represent the dielectric constants of the metal layer and the dielectric medium, and c is the velocity of light. To excite the SPPs at the metal-dielectric interface, the momentum i.e., the wave vector of exciting light, should be increased. Strong light absorption occurs because of the energy transfer during a reduction in the reflected light energy. This is the basic principle of the SPR technique.

\subsection{Dispersion Relationships}

Given its remarkable absorption features, gold is commonly utilized in the sensing fields within the structure of SPR sensors. The Drude-Lorentz model has been used instead of the basic Drude model to define the optical characteristic of the plasmonic layer of gold. Due to interband transition in gold, the drude model does not fit the gold dispersion relation in the 1.24-2.48 eV range. Using the Drude-Lorentz model defines the dielectric permittivity of gold as follows.\cite{PhysRevB.71.085416}
\begin{equation}\label{drude}
    \epsilon_{Au} = \epsilon_\infty - \frac{{\omega_D}^2}{\omega(\omega+j\gamma_D)}-\frac{\Delta\epsilon\Omega_L^2}{(\omega^2-\Omega_L^2)+j\Gamma_L\omega}
\end{equation}
Where $\epsilon_{Au}$ is the permittivity of the gold, $\epsilon_\infty$= 5.9673 is the permittivity of gold due to the bound electron response at high frequency. $\omega_D$ and $\gamma_D$ are the plasma frequency and damping frequency having values of  $\omega_D/2\pi$  =  2113.6 THz and $\gamma_D/2\pi$  = 15.92 THz. Weighting factor $\Delta\epsilon$ = 1.09, and the angular frequency can be expressed as $\omega = 2\pi c/\lambda$. Oscillator strength, $\Omega_L/2\pi$ = 650.07 THz and spectral-width of Lorentz oscillator, $\Gamma_L/2\pi$ = 104.86 THz. The refractive Index response of $Ta_2 O_5$ with frequency is evaluated from \cite{Bright2013InfraredOP} ($\lambda$ in micrometer here)
\begin{equation}\label{ta2O5}
    n_{Ta_2O_5}=2.06+\frac{0.025}{\lambda^2}  
\end{equation}

\subsection{Experimental Setup}
The practical sensing system with the proposed sensor is depicted in Fig\ref{fig:setup}. The light source is coupled to the PCF sensor via single-mode fiber (SMF), an interconnection link between two modules. Further, SMF connects the transmitted light to a high-resolution optical spectrum analyzer (OSA). Two navigation channels named 'in' and 'out' are used to control the flow of analyte into the sensor module. OSA visualizes power spectral density vs. operating frequency, and a computer connected to OSA analyzes the data to identify the analyte.  
\begin{center}
    \centering
    \includegraphics[width=0.8\textwidth]{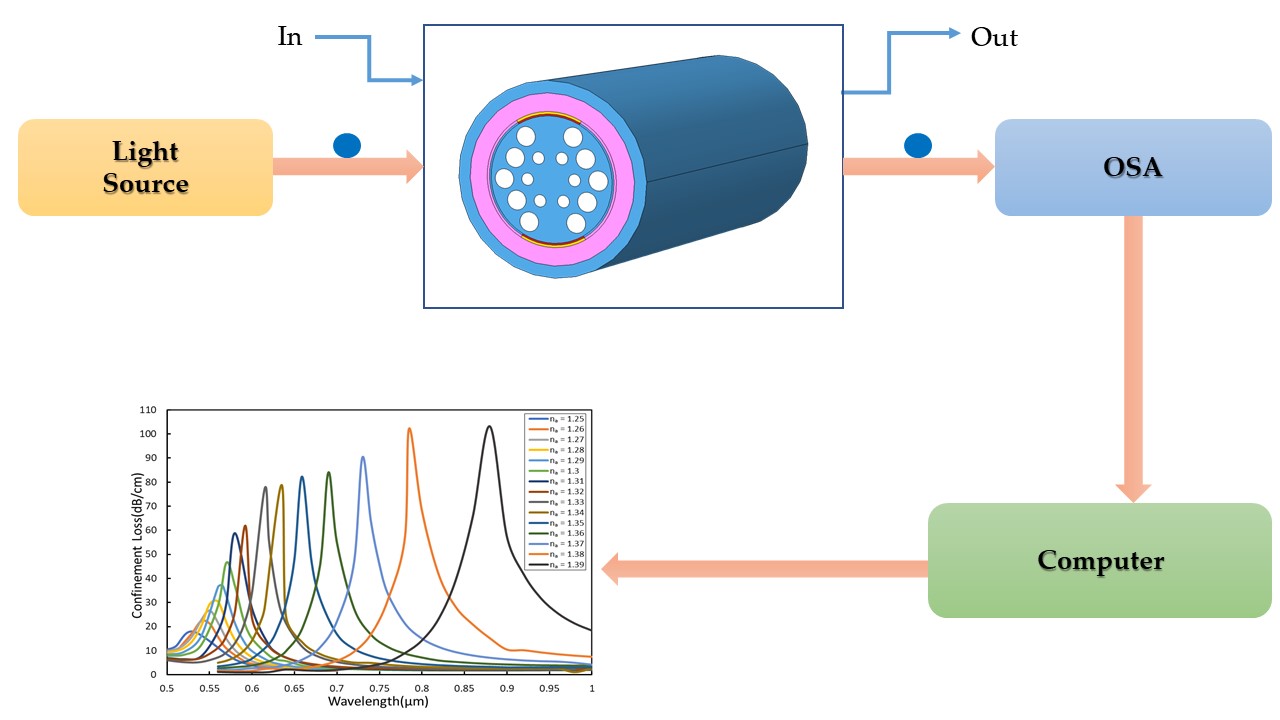}
    \captionof{figure}{Experimental setup diagram for practical sensing}
    \label{fig:setup}
\end{center}

\section{Performance Analysis}
\subsection{Confinement loss for varying analyte Refractive Index}
As the dispersion relation of SPP of the plasmonic metal-silica dielectric layer matches the evanescent wave k vector, consequently, excitation of the surface plasmons causes a significant energy transfer from the core to the cladding layer of the PCF. This reduction in energy from the core causes to couple of the core mode and surface plasmon mode resulting in a sharp confinement loss, calculated using the following equation. \cite{Shuai:12}
\begin{equation}\label{CL}
    \alpha_{CL}=8.686 \times k_o \times Im(n_{eff}) \times 10^4 dB/cm
\end{equation}
where $k_o =  2\pi/\lambda$  is the free space wave number. Fig.\ref{fig:gold_CL} and \ref{fig:ta2o5_CL} show the confinement loss spectra of gold film coating and $Ta_2 O_5$ overlay for the analyte range of 1.25 to 1.39. 

\subsection{Wavelength Sensitivity}
As the refractive index of the sample analyte changes, the overall effective index of the PCF varies, consequently varying the relative permittivity of the structure. The general behavior of permittivity as a function of frequency defines the resonance occurring wavelength at which the imaginary part of the permittivity is very high compared to the real part, causing absorption in the cladding region. Thus shifting the effective permittivity causes the shift of resonance wavelength at which maximum propagation loss occurs. Utilizing this change, the wavelength or spectral interrogation sensitivity (WIS) of the sensor can be obtained by following the equation 
\begin{equation}\label{WIS}
    S_{WIS} = \frac{\Delta\lambda_{peak}}{\Delta n_a} nm/RIU 
\end{equation}
Here, $\Delta\lambda_{peak}$ is the difference between two resonance wavelengths, and $\Delta n_a$ is the difference in related $n_a$.

\subsection{Amplitude Sensitivity}
Amplitude interrogation sensitivity (AIS) is another cost-effective performance parameter compared to the WIS method, as AIS can be measured at a single wavelength. AIS can be calculated using the equation \cite{6718021}
\begin{equation}\label{AIS}
    S_{AIS}(\lambda) = -\frac{1}{\alpha(\lambda,n_a)}\frac{\delta\alpha(\lambda,n_a)}{\delta n_a} {RIU}^{-1} 
\end{equation}
where $\alpha(\lambda,n_a)$ is the confinement loss of the core-driven mode of a specific analyte, $\delta\alpha(\lambda,n_a)$ is the difference between two adjacent loss curves at a specific wavelength and RI and $\delta n_a$ is known as the difference in analyte refractive index. Fig.\ref{fig: gold AS} and \ref{fig:ta2o5 AS} show the amplitude sensitivity of the proposed sensor with gold and $Ta_2 O_5$ coating for analyte RI 1.25 to 1.39.

\subsection{Resolution}
The sensor's resolution is defined as the ability to define the smallest change in the analyte's refractive index and is calculated by using the following formula \cite{article}
\begin{equation}\label{resolution}
    R = \frac{\Delta n_a \times \Delta\lambda_{min}}{\Delta\lambda_{peak}}
\end{equation}
where, $\Delta n_a$ = 0.01, $\Delta\lambda_{min}$ = 0.1 nm and $\Delta\lambda_{peak}$ is the difference between two adjacent resonance wavelengths.
\section{Result and Discussion}
To study the proposed PCF sensor performance, the design is engineered with two different plasmonic layers, viz. gold and $Ta_2 O_5$ deposited on gold separately. Wavelength interrogation sensitivity, amplitude interrogation sensitivity, and resolution are defined for all cases. All structures are TM polarized, and analysis is carried in visible to near IR range.
\subsection{Phase Matching and Mode Coupling}
SPR generation in the PCF structure works on the principle of maximum energy transfer between the core and spp mode at the resonance wavelength. The intersection of fundamental and spp mode dispersion relations numerically represents this resonant coupling phenomenon, implying equal propagation constants of the two modes.\\ 
Mode analysis was investigated by selecting electromagnetic waves, frequency-domain solver interface under the radio-frequency physics interface in COMSOL. 

\begin{figure}[!htb]
    \begin{subfigure}[b]{0.32\textwidth}
  \includegraphics[width=\linewidth]{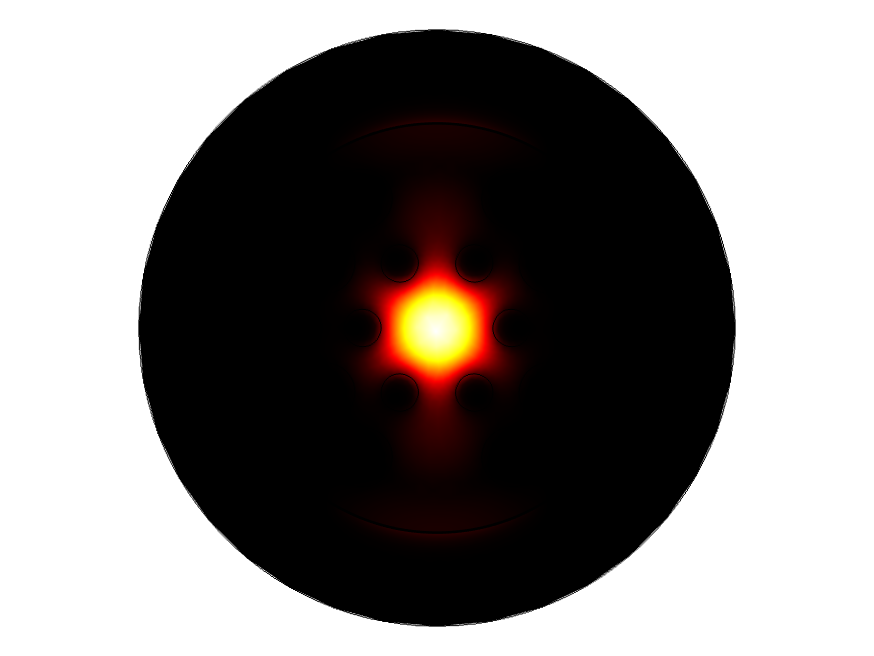}
  \caption{}
    \end{subfigure}
    \begin{subfigure}[b]{0.32\textwidth}
  \includegraphics[width=\linewidth]{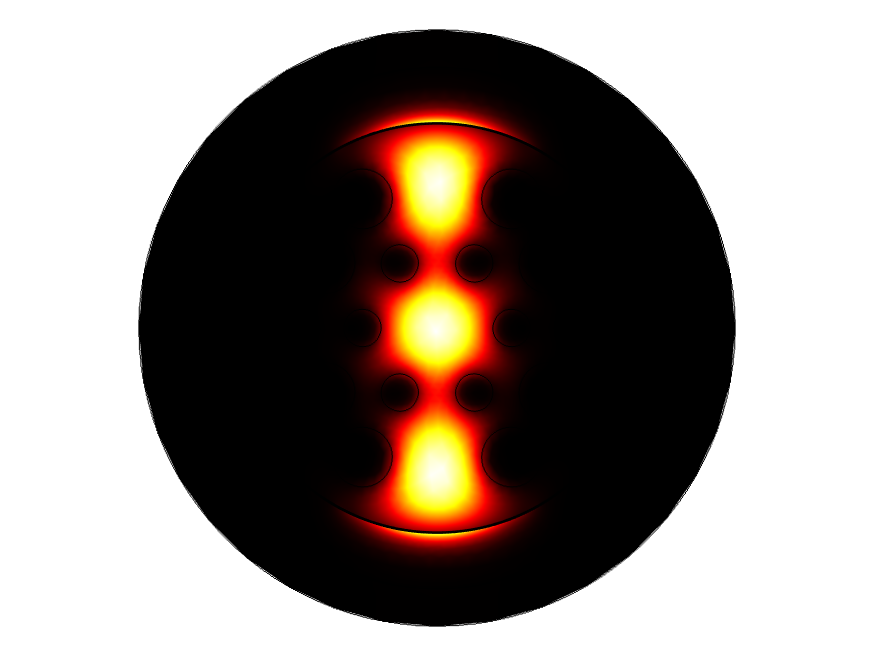}
  \caption{}
    \end{subfigure}
    \begin{subfigure}[b]{0.32\textwidth}
  \includegraphics[width=\linewidth]{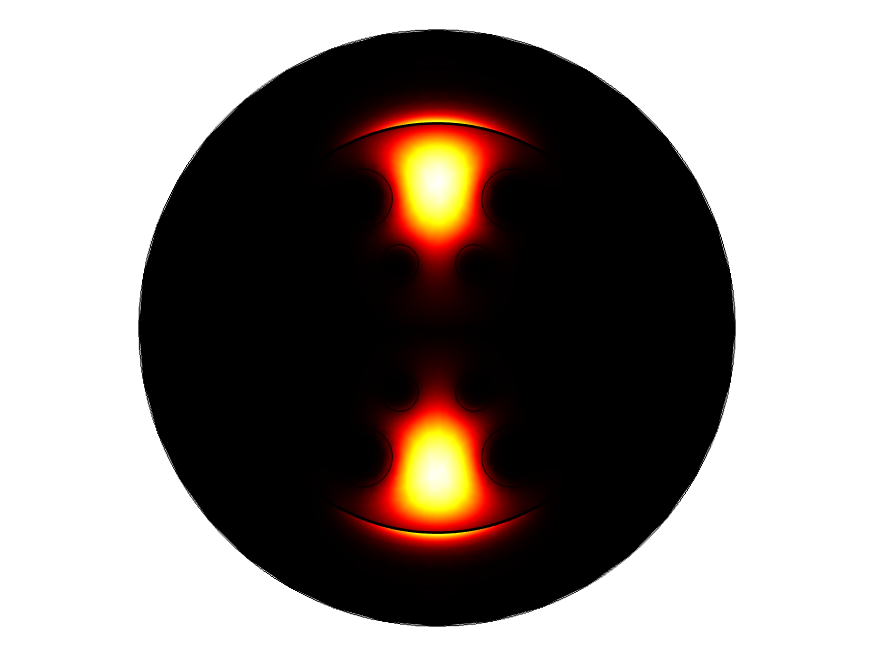}
  \caption{}
    \end{subfigure}
\caption{Optical field orientation of (a) core-driven mode at a random wavelength ($\lambda$ = 0.7 $\mu$m) (b) core-driven mode at the resonance wavelength ($\lambda$ = 0.8 $\mu$m) (c) plasmonic mode at resonance wavelength ($\lambda$ = 0.8 $\mu$m) using gold plasmonic film (when d = 1 $\mu$m, $d_1$ = 1.6 $\mu$m, $t_{Au}$ = 30 nm and $t_{ta_2O_5}$ = 10 nm for $n_a$ = 1.35)}\label{fig:core-spp modes}
\end{figure}
The optical field distribution for modes at different operating wavelengths for the proposed sensor is illustrated in Fig.\ref{fig:core-spp modes} for an analyte of RI 1.35. It should be noted that the x-polarization mode was considered at every wavelength as it shows more sensitivity compared to the y-polarization mode.
\vspace{2pt}
\begin{center}
    \centering
    \includegraphics[width=0.9\textwidth]{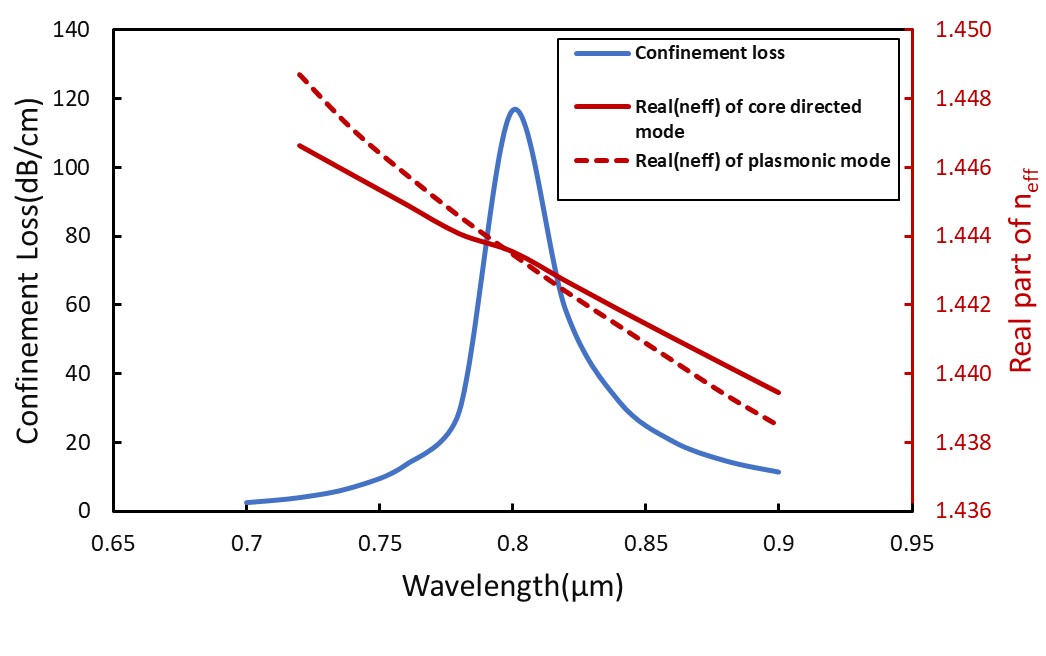}
    \captionof{figure}{Dispersion relation between the core-directed mode and plasmonic mode for gold plasmonic layer at $n_a$ = 1.35 (when d = 1 $\mu$m, $d_1$ = 1.6 $\mu$m, $t_{Au}$ = 30 nm and $t_{ta_2O_5}$ = 10 nm).}
    \label{fig:ta2o5_check}
\end{center}
In Fig.\ref{fig:ta2o5_check}, the dispersion curves for the above-mentioned modes are depicted for $n_a$ = 1.35, where the solid blue line indicates the confinement loss for the core mode. The real part of RI of both core and plasmonic mode is plotted as a function of wavelength. With increasing operating wavelength from 0.75 $\mu$m, the real part of the two modes approaches each other, and energy transfer from core to cladding also increases. At 0.8 $\mu$m of operating wavelength, the real portions of the effective RI of core and plasmonic mode percept each other resulting in a sharp peak of confinement loss. At this phase matching condition, maximum energy from the core leaks into the cladding region, causing surface plasmon resonance with a maximum loss of 117.04 dB/cm. 
\subsection{Optimization of Structural Parameters}
\begin{figure}[htbp]
     \centering
     \begin{subfigure}[b]{0.5\textwidth}
         \centering
         \includegraphics[width=\textwidth]{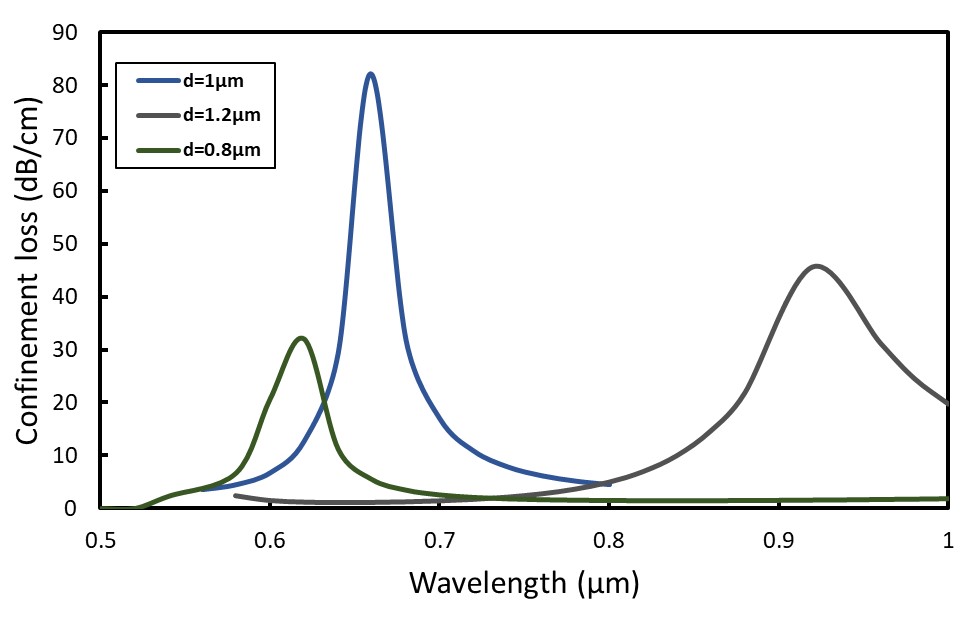}
         \caption{}
         \label{fig:small dia}
     \end{subfigure}
     \hfill
     \begin{subfigure}[b]{0.54\textwidth}
         \centering
         \includegraphics[width=\textwidth]{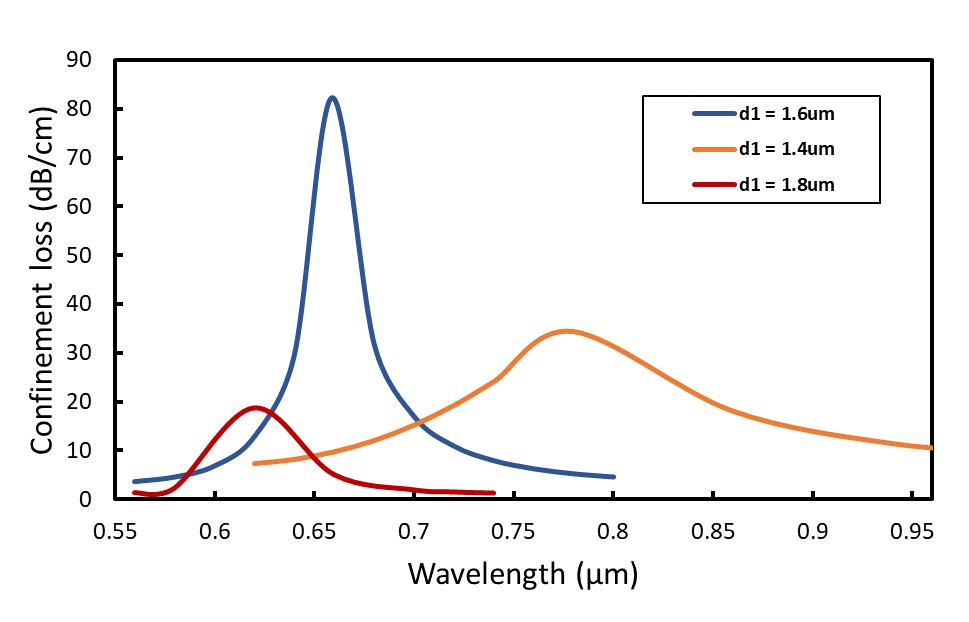}
         \caption{}
         \label{fig:large dia}
     \end{subfigure}
     \hfill
     \begin{subfigure}[b]{0.52\textwidth}
         \centering
         \includegraphics[width=\textwidth]{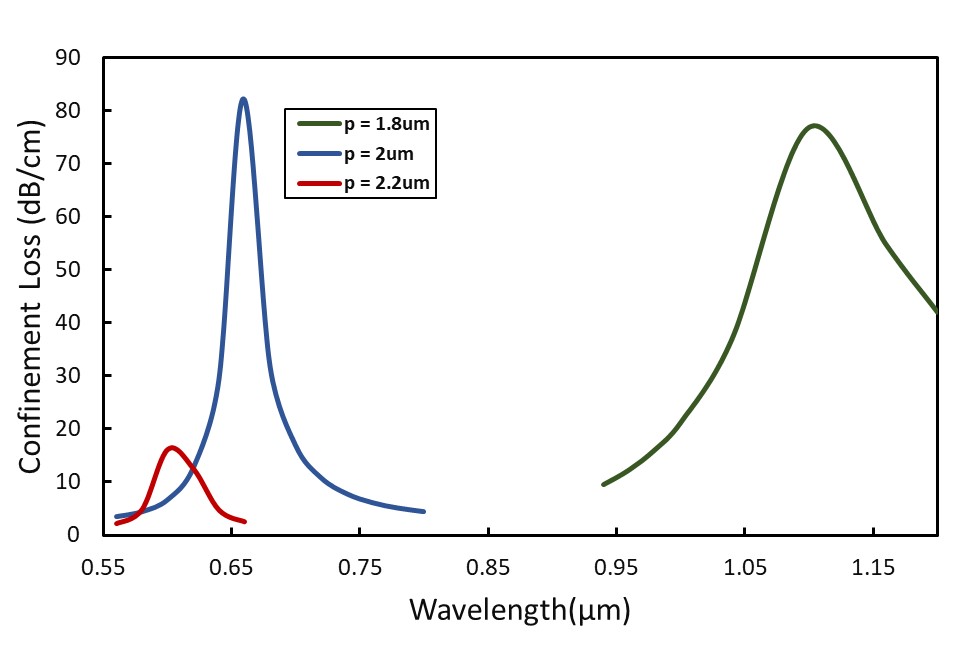}
         \caption{}
         \label{fig:pitch}
     \end{subfigure}
        \caption{Resonance spectra for (a) variation of smaller airholes diameter(d) (b) variation of larger air holes diameter($d_1$) (c) variation of pitch (p).  }
        \label{fig:optimization}
\end{figure}

\begin{figure}[htbp]
     \centering
     \begin{subfigure}[b]{0.5\textwidth}
         \centering
         \includegraphics[width=\textwidth]{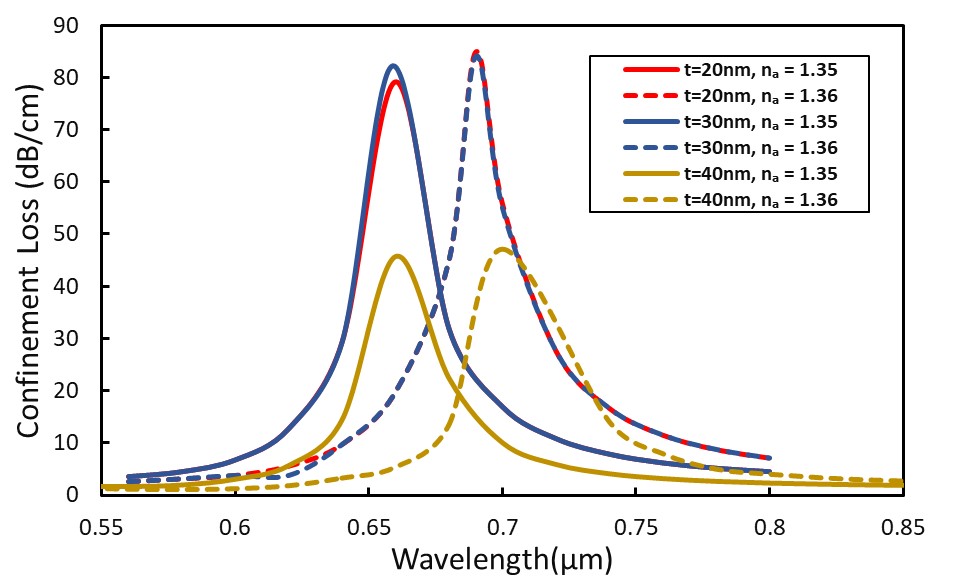}
         \caption{}
         \label{fig:gold thickness}
     \end{subfigure}
     \hfill
     \begin{subfigure}[b]{0.48\textwidth}
         \centering
         \includegraphics[width=\textwidth]{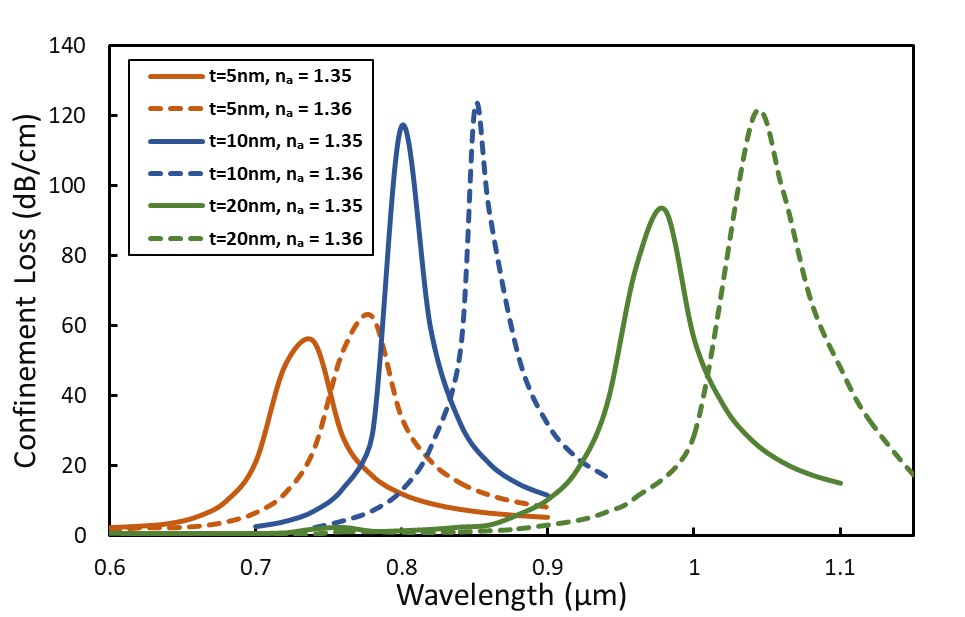}
         \caption{}
         \label{fig:ta2o5 thickness}
     \end{subfigure}
     \hfill
        \caption{Resonance spectra for (a) variation of the thickness of the gold layer (t) in the range of 20-40 nm (b) variation of the thickness of $ta_2O_5$ layer (t) in the range of 5-20 nm. }
        \label{fig:thickness}
\end{figure}

Variation of different structural parameters plays a crucial role while designing a photonic crystal fiber. One of the significant performance parameters for determining the effectiveness of a sensor is studying its dispersion profile. So, $\alpha(\lambda)$ for the variations of airhole diameters, pitch and plasmonic thickness is optimized in this section and represented in Fig.\ref{fig:optimization}. \\
Fig.\ref{fig:small dia} shows the loss spectrum when the smaller diameters ($d$) of the PCF structure is varied between 0.8-1.2 $\mu$m. During 0.8-1 $\mu$m, the loss at resonance increases as more light can be coupled through the missing air holes of 2nd loop. But it starts to gradually decrease with the increase of $d$ as larger inner holes tend to confine more light in the core. \\
Fig.\ref{fig:large dia} shows the effect of variation of outer airholes diameter ($d_1$) within the range of 1.4-1.8 $\mu$m. As the diameter increases from 1.4 $\mu$m to 1.6 $\mu$m, confinement loss blueshifts with an increment though it sharply decreases as the diameter increases. The loss curve peak is the most prominent with inner air hole diameter ($d$) = 1 $\mu$m and outer air hole diameter ($d_1$) = 1.6 $\mu$m.\\
Fig.\ref{fig:pitch} evidently shows the effect of pitch ($p$) or lattice constant on the sensor performance. Pitch is varied from 1.8-2.2 $\mu$m, and when p = 2 $\mu$m, the loss reaches its highest peak and gradually decreases after that as the $n_{eff}$ of the core mode decreases, appearing a blueshift. \\
The plasmonic layer is used to enhance the coupling between spp and core mode at the resonance wavelength and the maximum loss is highly affected by the optimum thickness of the coating layer. Fig.\ref{fig:thickness} shows the loss spectra for different thicknesses of plasmonic layers for the RI of 1.35 and 1.36. Due to its greater surface plasmon field, a thicker gold layer allows the penetration of evanescent field causing a better core and spp coupling, which is visible for the gold thickness of 20 nm and 30 nm in Fig.\ref{fig:gold thickness}. After a certain value of thickness, red shifting of the loss curve is noticeable. This phenomenon occurs as a larger gold layer requires an incident evanescent field of a higher wavelength to penetrate the thickness. The highest loss peak is found with 20 nm gold layer on the PCF. \\
Fig.\ref{fig:ta2o5 thickness} depicts the effect of thickness of $Ta_2O_5$ coating on the gold layer to enhance the mode coupling. Evidently, due to its higher refractive index, a thicker $Ta_2O_5$ requires more energy i.e. larger wavelength, to penetrate the plasmonic layer, causing an increase in loss peak. This is the case shown for $Ta_2O_5$ thickness of 5 nm and 10 nm. Further increasing the $Ta_2O_5$ thickness redshifts the loss peaks towards the IR region and the maximum loss also decreases gradually. Table \ref{Table 1} summarizes the detailed observation of resonance wavelength and loss peak for varying different structural parameters.
\begin{table}
\caption{Resonance wavelength and losses for varying structural parameters thickness for RI of 1.35}
\label{Table 1}
\centering
\tiny
\begin{tabular}{llll}
\hline
\multicolumn{2}{l}{\textbf{Structural   parameter Thickness ($\mu$m) }}       & \textbf{Resonance Wavelength ($\mu$m)} & \textbf{Loss (dB/cm)} \\
\hline
\multirow{3}{*}{$d$ ($d_1$ = 1.6 $\mu$m, p = 2 $\mu$m)} & 0.8 & 0.62                      & 32.03        \\
                                           & 1   & 0.659                     & 82.23        \\
                                           & 1.2 & 0.92                      & 45.72        \\
                                           \hline
\multirow{3}{*}{$d_1$ ($d$ = 1 $\mu$m, p = 2 $\mu$m)}      & 1.4 & 0.78                      & 34.48        \\
                                           & 1.6 & 0.659                     & 82.23        \\
                                           & 1.8 & 0.62                      & 18.64        \\
                                           \hline
\multirow{3}{*}{$p$ ($d$ = 1 um, $d_1$ = 1.6 $
\mu$m)} & 1.8 & 1.1                       & 76.93        \\
                                           & 2   & 0.659                     & 82.23        \\
                                           & 2.2 & 0.6                       & 16.17        \\
                                           \hline
\multirow{3}{*}{$t_{Au}$}                     & 20  & 0.66                      & 79.13        \\
                                           & 30  & 0.659                     & 82.23        \\
                                           & 40  & 0.66                      & 45.76        \\
                                           \hline
\multirow{3}{*}{$t_{Ta_2O_5}$}                  & 5   & 0.74                      & 55.35        \\
                                           & 10  & 0.8                       & 117.04       \\
                                           & 20  & 0.98                      & 92.96     \\
\hline                                           
\end{tabular}
\end{table}

\subsection{Analysis with gold layer}
Incremental analyte RI retrenches the disparity between core and plasmonic mode effective index, resulting in the evanescent field pairing with gold, consequently increasing the confinement loss peak. For the analyte RI range of 1.25 to 1.39 with the step of 0.01, the loss profile is shown in Fig.\ref{fig:gold_CL}. Confinement losses are calculated using equation \ref{CL}.
\begin{center}
    \centering
    \includegraphics[width=0.7\textwidth]{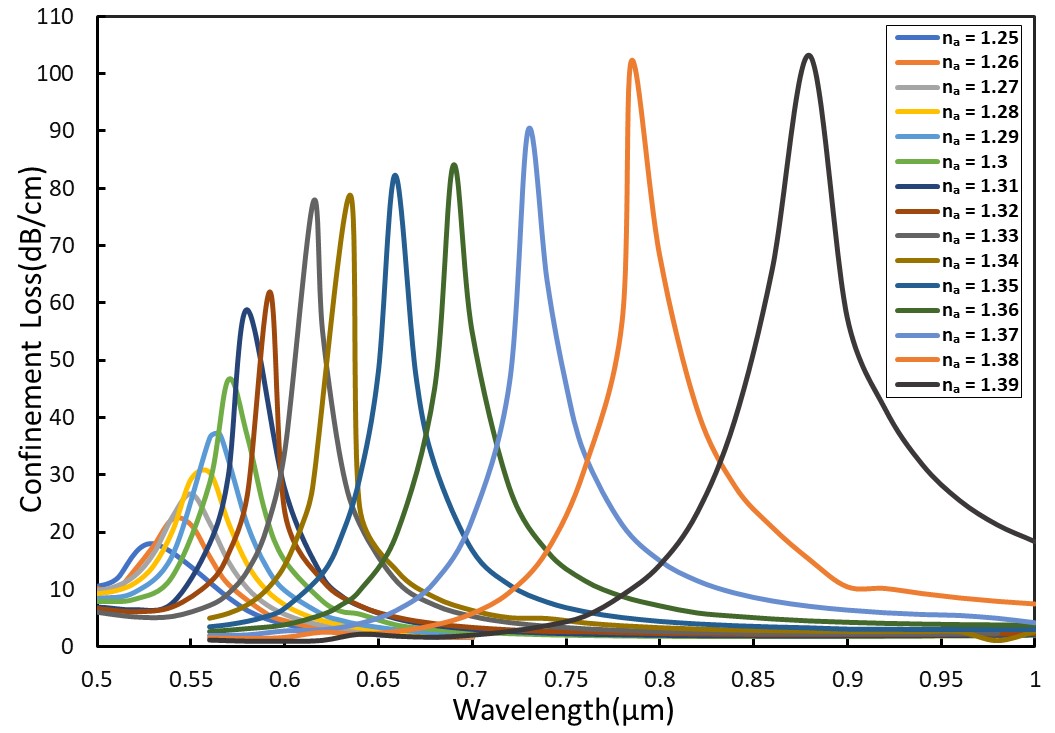}
    \captionof{figure}{Dispersion profile for variation in analyte RI (a) between 1.25-1.39 for gold film}
    \label{fig:gold_CL}
\end{center}
From the red shifting of resonance wavelength due to kinetic energy binding and propagation constant of higher RI analyte, maximum WIS is found 9500 nm/RIU for an analyte RI of 1.38 using equation \ref{WIS}.

\begin{figure}[htbp]
     \centering
     \begin{subfigure}[b]{0.55\textwidth}
         \centering
         \includegraphics[width=\textwidth]{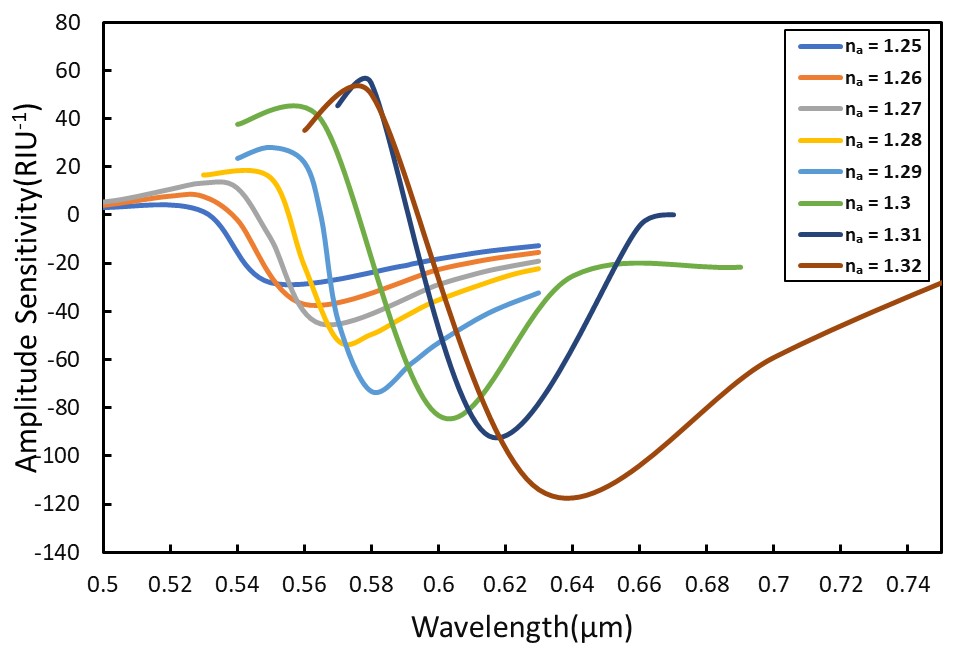}
         \caption{}
         \label{fig:gold AS1}
     \end{subfigure}
     \hfill
     \begin{subfigure}[b]{0.53\textwidth}
         \centering
         \includegraphics[width=\textwidth]{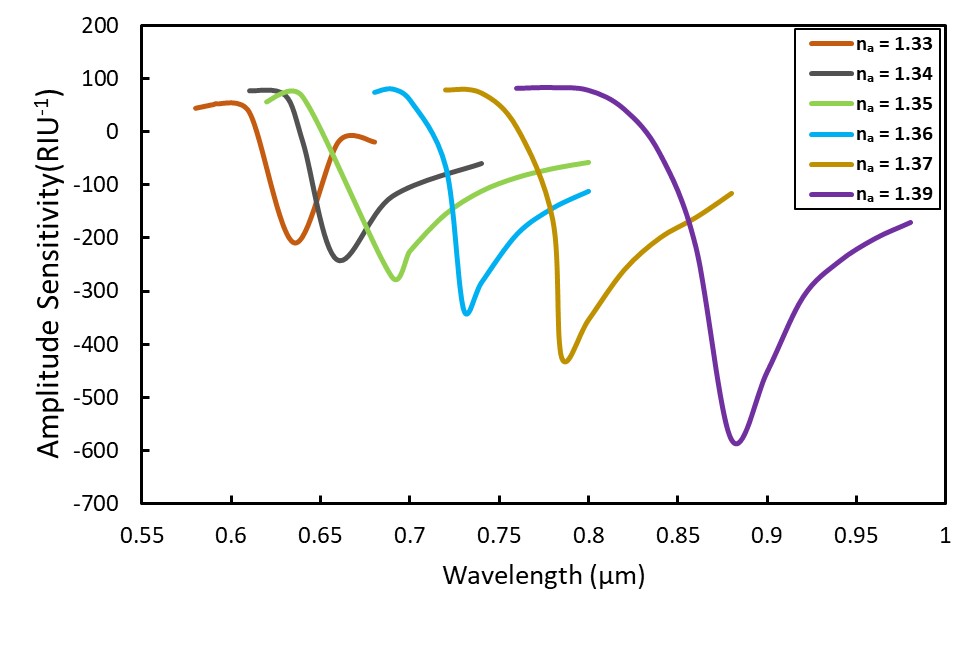}
         \caption{}
         \label{fig:gold AS2}
     \end{subfigure}
     \hfill
        \caption{AIS profile for variation in analyte RI (a) between 1.25-1.32 (b)between 1.33-1.38 for gold film. }
        \label{fig: gold AS}
\end{figure}

Fig.\ref{fig: gold AS} shows the amplitude or intensity interrogation sensitivity profile that is based on the change of the power at a fixed wavelength near the SPR peak for varying RI and is calculated from equation \ref{AIS}.
The resolution of the sensor is calculated from equation \ref{resolution}. The minimum resolution of this sensor with gold plasmonic layer has been found $1.05\times 10^{-5}$ RIU. This means any change in the refractive index of the analyte between every 0.01 can be detected precisely by the proposed sensor with gold plasmonic layer. 

\subsection{Analysis with $Ta_2O_5$ coating on gold layer}
\begin{figure}[htbp]
     \centering
     \begin{subfigure}[b]{0.55\textwidth}
         \centering
         \includegraphics[width=\textwidth]{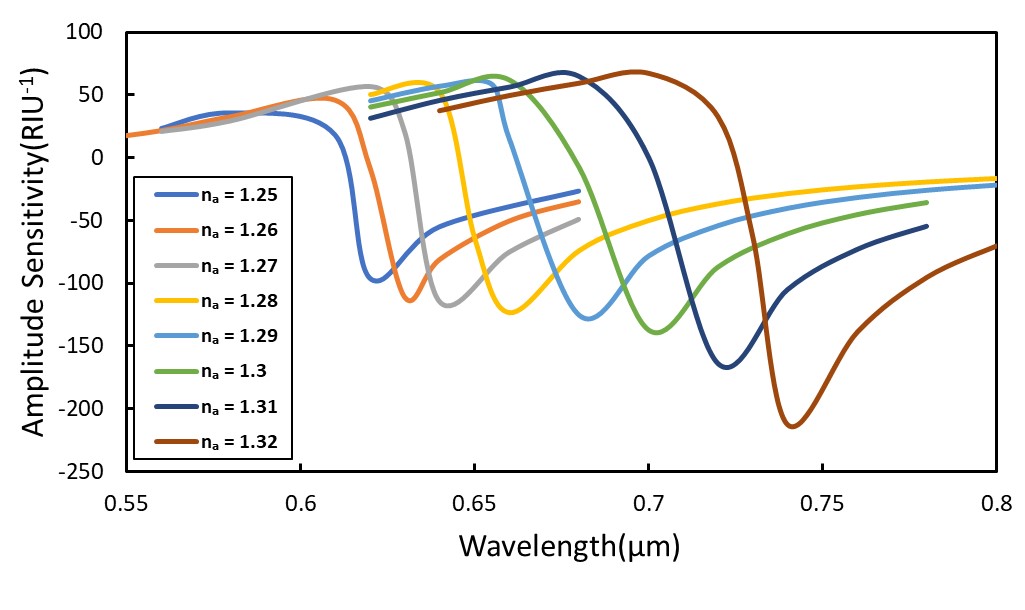}
         \caption{}
         \label{fig:ta2o5 AS1}
     \end{subfigure}
     \hfill
     \begin{subfigure}[b]{0.53\textwidth}
         \centering
         \includegraphics[width=\textwidth]{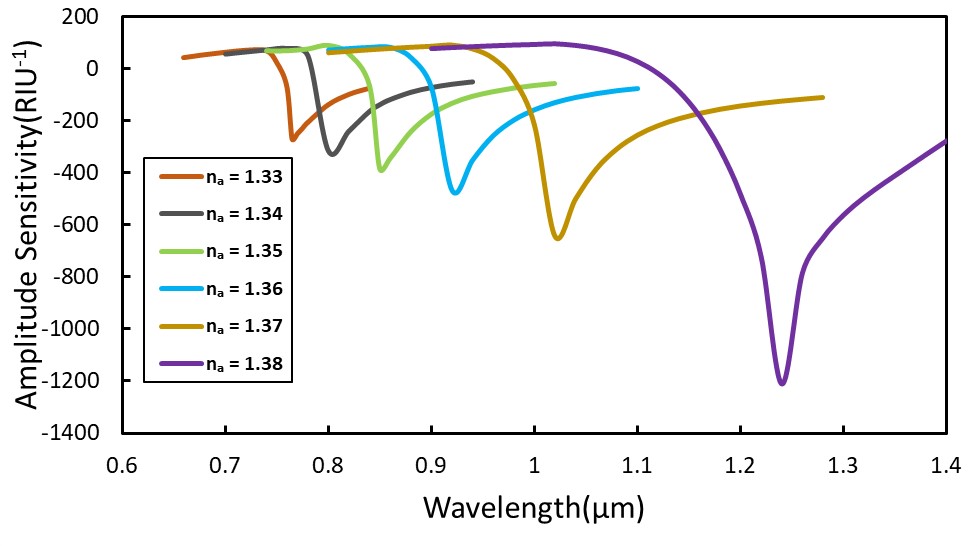}
         \caption{}
         \label{fig:ta2o5 AS2}
     \end{subfigure}
     \hfill
        \caption{AIS profile for variation in analyte RI (a) between 1.25-1.32 (b)between 1.33-1.38 for $Ta_2O_5$-coated gold film. }
        \label{fig:ta2o5 AS}
\end{figure}

High waveguide loss of 1 dB/cm at the infrared region makes $Ta_2 O_5$ depositing on the gold layer justified by benefiting the energy escaping from the core to the metal-coating layer. An extra-thin film of $Ta_2 O_5$  ( $t_{Ta_2 O_5}$= 10 nm) is used to ensure better bonding of gold with glass substrates as required in the practical fabrication of gold-covered biosensors.
To study the performance of deposition of $Ta_2 O_5$, all the optical characteristic parameters are again calculated using equations \ref{CL}, \ref{WIS}, \ref{AIS}, and \ref{resolution}.
\begin{center}
    \centering
    \includegraphics[width=0.7\textwidth]{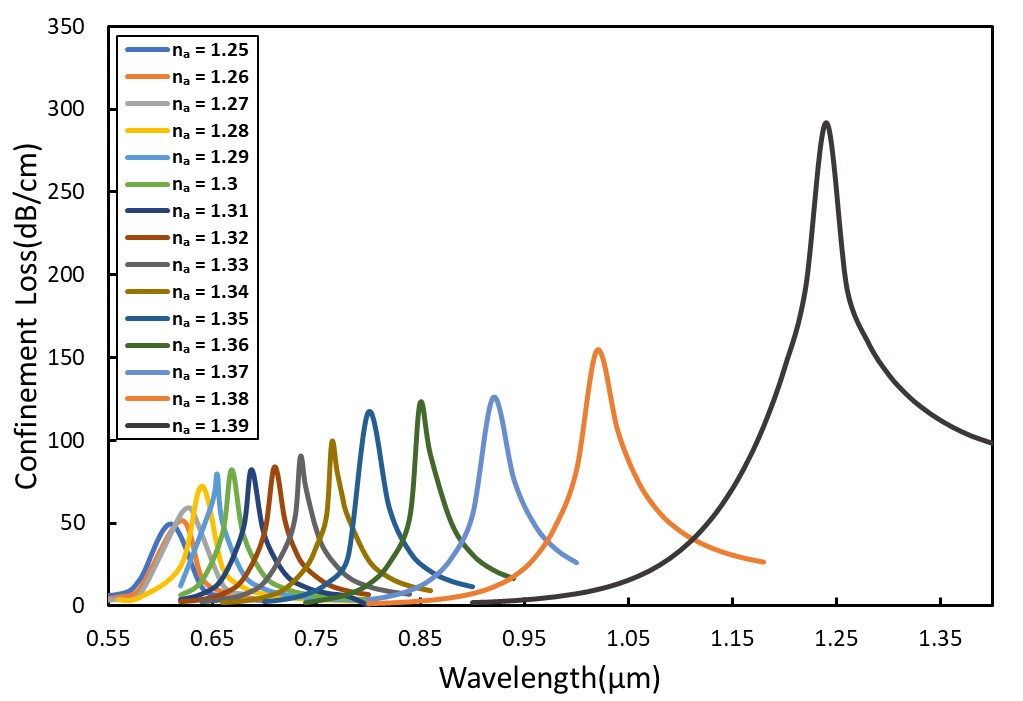}
    \captionof{figure}{Dispersion profile for variation in analyte RI (a) between 1.25-1.39 for $Ta_2O_5$-coated gold film}
    \label{fig:ta2o5_CL}
\end{center}
The dispersion profile of $Ta_2 O_5$ depicted in Fig.\ref{fig:ta2o5_CL} shows a more significant shift of resonance wavelength comparing the response without the $Ta_2 O_5$ film. Deposition of this metal oxide surface on gold contemplated to form durable adhesion between gold-silica surface while having low electromagnetic energy consumption.\\
Operating wavelength after depositing $Ta_2 O_5$ shifted from visible to near-infrared region, the spectral coverage can be estimated between 550 nm and 1400 nm. Fig.\ref{fig:ta2o5 AS} shows the AIS profiles for the RI range of 1.25 to 1.38.
\newpage

\begin{table}
\caption{Performance analysis of gold film and $Ta_2O_5$ coating on gold of the proposed sensor (with $d$ = 1 $\mu$m, $d_1$ = 1.6 $\mu$m and $p$ = 2 $\mu$m) between the analyte range of 1.25-1.39}
\label{Table 2}
\centering
\tiny
\begin{longtblr}
[
  label = none,
  entry = none,
]{
  width = \linewidth,
  colspec = {Q[60]Q[58]Q[113]Q[63]Q[113]Q[63]Q[113]Q[58]Q[112]Q[65]Q[112]},
  cells = {c},
  cell{1}{1} = {r=2}{},
  cell{1}{2} = {c=2}{0.15\linewidth},
  cell{1}{4} = {c=2}{0.15\linewidth},
  cell{1}{6} = {c=2}{0.15\linewidth},
  cell{1}{8} = {c=2}{0.12\linewidth},
  cell{1}{10} = {c=2}{0.25\linewidth},
  hline{1,3} = {-}{},
  hline{2} = {2-11}{},
  hline{18} = {-}{}
}
\textbf{RI} & \textbf{Resonance Wavelength} &                                             & \textbf{CL Peak (dB/cm)} &                                             & \textbf{AIS ($RIU^{-1}$)} &                                             & \textbf{WIS(nm/RIU)} &                                            & \textbf{Resolution(RIU)} &                                            \\
                    & \textbf{Gold (Au)}            & \textbf{Gold (Au) with $Ta_2O_5$ Layer} & \textbf{Gold (Au)}       & \textbf{Gold (Au) with $Ta_2O_5$ Layer} & \textbf{Gold (Au)}             & \textbf{Gold (Au) with $Ta_2O_5$ Layer} & \textbf{Gold (Au)}   & \textbf{Gold(Au) with $Ta_2O_5$ Layer} & \textbf{Gold (Au)}  & \textbf{Gold(Au) with $Ta_2O_5$ Layer} \\
1.25                & 0.53                          & 0.61                                        & 17.95              & 49.46                                 & 28.05                    & 95.93                                 & 1000                 & 1000                                       & 1$\times10^{-4}$          & 1$\times10^{-4}$                                 \\
1.26                & 0.54                          & 0.62                                        & 22.36              & 51.49                                   & 37.15                    & 111.49                                 & 1000                 & 1000                                       & 1$\times10^{-4}$          & 1$\times10^{-4}$                                 \\
1.27                & 0.55                          & 0.63                                        & 26.70              & 58.16                                 & 45.09                    & 134.57                                 & 1000                 & 1000                                       & 1$\times10^{-4}$          & 1$\times10^{-4}$                                 \\
1.28                & 0.56                          & 0.64                                        & 30.23              & 72.26                                 & 51.77                    & 123.91                                      & 500                  & 1550                                       & 2$\times10^{-4}$          & 6.45$\times10^{-5}$                            \\
1.29                & 0.565                         & 0.6555                                      & 37.23              & 79.71                                 & 73.02                    & 125.33                                     & 500                  & 1350                                       & 2$\times10^{-4}$          & 7.41$\times10^{-5}$                           \\
1.3                 & 0.57                          & 0.669                                       & 37.23              & 82.12                                  & 83.21                    & 137.51                                     & 1000                 & 1900                                       & 1$\times10^{-4}$          & 5.26$\times10^{-5}$                            \\
1.31                & 0.58                          & 0.688                                       & 58.76              & 82.20                                 & 91.80                    & 164.19                                     & 1200                 & 2200                                       & 8.33$\times10^{-5}$       & 4.55$\times10^{-5}$                            \\
1.32                & 0.592                         & 0.71                                        & 61.88              & 83.74                                 & 113.84                    & 213.03                                     & 2300                 & 2500                                       & 4.35$\times10^{-5}$     & 4$\times10^{-5}$                                 \\
1.33                & 0.615                         & 0.735                                       & 77.62              & 90.28                                  & 208.18                    & 270.66                                     & 2000                 & 3000                                       & 5$\times10^{-5}$          & 3.33$\times10^{-5}$                            \\
1.34                & 0.635                         & 0.765                                       & 78.87              & 98.90                                 & 240.68                    & 317.61                                     & 2400                 & 3500                                       & 4.17$\times10^{-5}$      & 2.86$\times10^{-5}$                            \\
1.35                & 0.659                         & 0.8                                         & 82.25              & 117.04                                 & 272.81                    & 344.92                                     & 3100                 & 5000                                       & 3.23$\times10^{-5}$        & 2$\times10^{-5}$                                \\
1.36                & 0.69                          & 0.85                                        & 84.07              & 122.59                                 & 336.09                    & 470.46                                     & 4000                 & 7000                                       & 2.5$\times10^{-5}$        & 1.43$\times10^{-5}$                            \\
1.37                & 0.73                          & 0.92                                        & 90.33              & 125.71                                 & 426.72                     & 643.01                                 & 5500                 & 10000                                      & 1.82$\times10^{-5}$     & 1$\times10^{-5}$                                 \\
1.38                & 0.785                         & 1.02                                        & 102.31              & 154.45                                 & 579.26                    & 1209.21                                     & 9500                 & 22000                                      & 1.05$\times10^{-5}$     & 4.55$\times10^{-6}$                            \\
1.39                & 0.88                          & 1.24                                        & 103.20              & 291.79                                  & $\hyp{}$                              & $\hyp{}$                                           & $\hyp{}$                    & $\hyp{}$                                         & $\hyp{}$                 & $\hyp{}$          \\
\end{longtblr}
\end{table}

Looking closely at Table \ref{Table 2}  for the various parameter values obtained in the intermediate portion, we can see that these parameter values differ from those obtained for the same parameters without any metal-oxide deposition. $Ta_2 O_5$ layer caused an improvement in the original AIS (with only gold, which is about 579.26 $RIU^{-1}$) to 1209.21 $RIU^{-1}$. Two main factors play vital roles in this. In the case of a metallic (i.e., gold) adhesion layer, high surface absorption minimizes the intensity of SPP. In contrast, a dielectric layer reduces the intensity of SPP due to its refractive index. Overall, the AIS is retrenched with a metal-oxide layer.\\
The linearity response of the sensor develops a relationship between two variables where resonance wavelength is plotted as a function of analyte RI. Fig.\ref{fig: linearity} depicts the linear response of the proposed sensor for both gold film and $Ta_2O_5$ coated gold film. The regression equations for gold, gold coated $Ta_2 O_5$ are y = 2.0668x - 2.0961 and y = 3.4902x - 3.8369, respectively, where x is the analyte RI and y denotes the resonance wavelength. The values of $R^2$ are found to be 0.8404 and 0.7883, which are signs of good linearity.  
\begin{center}
    \centering
    \includegraphics[width=0.65\textwidth]{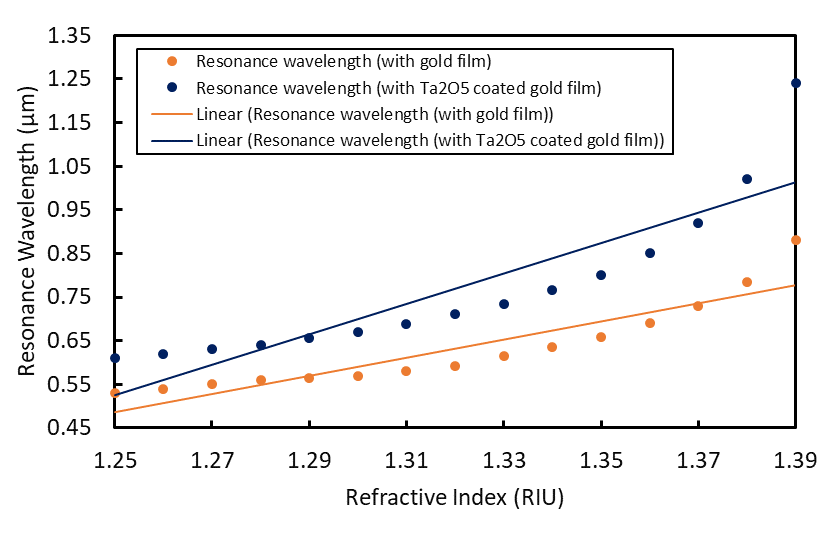}
    \captionof{figure}{Linearity response of reported sensor with gold and $Ta_2O_5$ coated gold film}
    \label{fig: linearity}
\end{center}

\section{Comparison with existing documented sensors}
Table \ref{Table 3} presents a detailed comparison of our proposed sensor based on its WIS, AIS, resolution, and the sensing range of analyte RI. Characterization parameters for both gold and $Ta_2O_5$ coated gold film were considered. This comparison depicts that our proposed sensor outperforms existing works having better AIS, WIS as well as a broader sensing range.
\newpage
\begin{table}
\caption{Performance discrimination of the proposed arc-shaped sensor with existing works}
\label{Table 3}
\centering 
\tiny
\begin{longtblr}
[
  label = none,
  entry = none,
]{
  width = \linewidth,
  colspec = {Q[60]Q[58]Q[113]Q[63]Q[113]Q[63]Q[113]Q[58]},
  hline{1,2} = {-}{},
  hline{13} = {-}{},
}
Ref                           & PCF Structure                & Plasmonic Material                         & Maximum WIS (nm/RIU) & AIS ($RIU^{-1}$) & Resolution (RIU)                & RI Range     \\
\cite{7361738}                      & Circular                     & Cupper with Graphene Layer                 & 2000               & 140                          & 5$\times10^{-5}$                        & 1.33-137     \\
\cite{Luan:15}                     & D-shape with hollow core     & Gold                                       & 3700               & $\hyp{}$                            & 2.7$\times10^{-5}$                     & 1.33-1.37    \\
\cite{Kravets2014GrapheneprotectedCA}                       & Exposed Core Structure       & Gold with Graphene                         & 2290               & $\hyp{}$                            & $\hyp{}$                               & 1.333-1.3688 \\
\cite{s21030818}                     & D-shape PCF                  & Gold with Graphene                         & 4200               & 450                          & 2.3$\times10^{-5}$                     & 1.32-1.41    \\
\cite{SHAFKAT2021166418}                      & Circular shape               & Different Noble Materials and their Alloys & 26000              & 149.47                       & 3.84$\times10^{-6}$                    & 1.37-1.40    \\
\cite{DAS2021100904}                       & D-shape PCF                  & Silver with  $Ta_2O_5$                     & 50000              & $\hyp{}$                            & $\hyp{}$                               & 1.38-1.42    \\
\cite{unknown}                      & Dual-Core                    & Tantalum                                   & 10400              & $\hyp{}$                            & 9.615$\times10^{-6}$                   & 1.25-1.39    \\
\cite{ISLAM2023106266}                      & Open   Channel-Based PCF-SPR & Gold                                       & 7000               & 593.61                       & 1.43$\times10^{-5}$                     & 1.33-1.40    \\
\cite{Melwin2023}                      & Arc Shaped                   & Gold                                       & 14100              & 109                          & 7.09$\times10^{-6}$                     & 1.32-1.37    \\
\setcellgapes{2}{}{This   work} & Arc Shape                    & Gold                                       & 9500               & 1729.55                      & 1.06$\times10^{-5}$ & 1.25-1.39    \\
                              & Arc Shape                    & Gold with $Ta_2O_5$                        & 22000              & 1668.23                      & 4.54$\times10^{-6}$ & 1.25-1.39   
\end{longtblr}
\end{table}

\section{Fabrication challenges}
The stack-and-draw method is proposed as a fabrication technique to address the challenge of designing PCFs with circular air holes \cite{8736236}. This method utilizes a combination of thin wall capillaries, thick wall capillaries, and solid rods to create the desired PCF structure while accommodating circular air holes. Using different types of capillaries and solid rods makes it possible to achieve the desired diameter variations required for the air holes in the PCF.
The stack-and-draw method involves stacking the capillaries and solid rods in a specific order to create a preform. The preform is then heated and drawn down to the desired PCF size. The thin wall capillaries are used for larger-diameter air holes, which can easily collapse during the drawing process. On the other hand, thick wall capillaries are used for smaller-diameter air holes to maintain their shape during drawing. Solid rods are used to fill the void areas between the air holes. Implementing the stack-and-draw technique allows for precise control over the dimensions of the air holes in the PCF. It provides a flexible and efficient approach to fabricating PCFs with circular air holes, overcoming the challenges associated with their design.
\begin{center}
    \centering
    \includegraphics[width=0.5\textwidth]{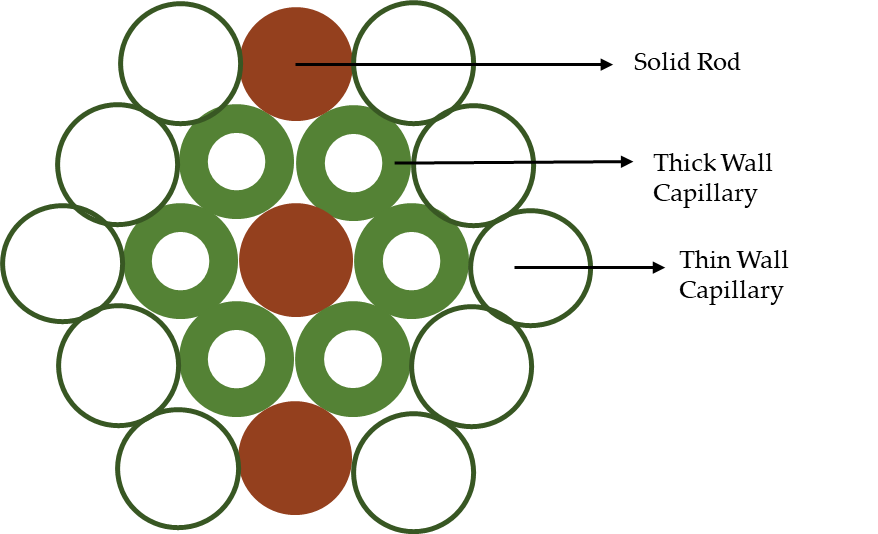}
    \captionof{figure}{The stacked mount of the proposed sensor.}
    \label{fig: fabrication}
\end{center}
In Fig.\ref{fig: fabrication}, the stacked mount of the proposed sensor is depicted. To create a polished and uniform metal coating on the outer surface of the fiber structure, techniques such as wheel polishing, chemical vapor deposition (CVD), and atomic layer deposition can be employed \cite{doi:10.1080/01468030.2013.879680}. While PCF-SPR sensors show great potential, most existing work has been limited to theoretical demonstrations rather than practical fabrication. However, theoretical statements need to be supported by practical production. Various manufacturing techniques are available for PCF fabrication, including extrusion, stack-and-draw, 3-D printing, and sol-gel casting \cite{RIFAT2017311},\cite{8736236}. The stack-and-draw method may not be feasible for PCFs with non-circular air holes, and alternative methods like extrusion and 3-D printing can be implemented for fabrication.\\
As coating processes like thermal evaporation or radio-frequency sputtering can produce surface roughness that affects sensing properties, it is crucial to achieve uniform thickness of the plasmonic material coating on the inner surface of the PCF \cite{Haider:18},\cite{photonics4010018}. The complex process of chemical vapor deposition (CVD) offers a way to reduce roughness. Techniques like focused ion-beam micromachining or femtosecond laser-based methods, can be used in PCF-SPR sensor structures that need selective analyte infiltration [48]. Repositioning the PCF from its setup, however, might affect coupling and lead to unreliable results \cite{RIFAT2017311}. This problem can be solved by creating an external sensing mechanism based on a PCF-SPR sensor model. Cost-effective PCF-SPR sensor models for biochemical identification across a wide refractive index range will likely be made possible by improvements in fabrication technologies.

\section{Conclusion}
In this work, an arc-shaped PCF-based SPR sensor with gold coating and $Ta_2 O_5$ overlay on gold has been proposed, and its performance has been analyzed using the Finite Element Method (FEM). By thoroughly analyzing the shift in the peak resonance 
wavelength, the proposed sensor exhibits a maximum wavelength sensitivity of 22000 nm/RIU with the resolution of $4.54\times10^{-6}$ RIU within the analyte range of 1.25 to 1.39 when a $Ta_2 O_5$ was coated on gold layer. The deposition of $Ta_2 O_5$ coating over gold enhances performance, resulting in a remarkable increase in the Wavelength Interrogation Sensitivity (WIS), while maintaining the same sensitivity range.  \\
This study utilized $Ta_2 O_5$ as the adhesive layer to effectively support the plasmonic material coating of gold. To ensure proper bonding of gold with the glass substrate in the fabrication of gold-covered plasmonic biosensors, it is crucial to include an extra-thin film of metal/dielectric. In the future, adhesive layers made of metals such as Ti, Cr, Mo, Ni, and Zn can be utilized to optimize better sensor performance.

\newpage
 \bibliographystyle{elsarticle-num} 
 \bibliography{cas-refs}




\end{document}